\documentclass[pra,twocolumn,showpacs,preprintnumbers,amsmath,amssymb]{revtex4}
\usepackage{amsfonts}
\usepackage{amsmath}
\usepackage{amssymb}
\usepackage{epsfig}
\usepackage{graphicx}%
\ifx\pdfoutput\relax\let\pdfoutput=\undefined\fi
\newcount\msipdfoutput
\ifx\pdfoutput\undefined\else
\ifcase\pdfoutput\else
\ifx\paperwidth\undefined\else
\ifdim\paperheight=0pt\relax\else\pdfpageheight\paperheight\fi
\ifdim\paperwidth=0pt\relax\else\pdfpagewidth\paperwidth\fi
\fi\fi\fi
\begin{document}
\title{Experimental test of the tradeoff relation in weak measurement}

\author{Geng Chen}
\affiliation{Key Laboratory of Quantum Information, University of
Science and Technology of China, CAS, Hefei, 230026, China}

\author{Yang Zou}
\affiliation{Key Laboratory of Quantum Information, University of
Science and Technology of China, CAS, Hefei, 230026, China}

\author{Xiao-Ye Xu}
\affiliation{Key Laboratory of Quantum Information, University of
Science and Technology of China, CAS, Hefei, 230026, China}

\author{Jian-Shun Tang}
\affiliation{Key Laboratory of Quantum Information, University of
Science and Technology of China, CAS, Hefei, 230026, China}

\author{Yu-Long Li}
\affiliation{Key Laboratory of Quantum Information, University of
Science and Technology of China, CAS, Hefei, 230026, China}

\author{Yong-Jian Han$\footnote{email:smhan@ustc.edu.cn}$}
\affiliation{Key Laboratory of Quantum Information, University of
Science and Technology of China, CAS, Hefei, 230026, China}

\author{Chuan-Feng Li$\footnote{email:cfli@ustc.edu.cn}$}
\affiliation{Key Laboratory of Quantum Information, University of
Science and Technology of China, CAS, Hefei, 230026, China}

\author{Guang-Can Guo}
\affiliation{Key Laboratory of Quantum Information, University of
Science and Technology of China, CAS, Hefei, 230026, China}

\author{Hai-qiao Ni}
\affiliation{National Laboratory for Superlattices and Microstructures, Institute of Semiconductors, Chinese Academy of Sciences, P.O. Box 912, Beijing 100083, China}

\author{Ying Yu}
\affiliation{National Laboratory for Superlattices and Microstructures, Institute of Semiconductors, Chinese Academy of Sciences, P.O. Box 912, Beijing 100083, China}

\author{Mi-feng Li}
\affiliation{National Laboratory for Superlattices and Microstructures, Institute of Semiconductors, Chinese Academy of Sciences, P.O. Box 912, Beijing 100083, China}

\author{Guo-wei Zha}
\affiliation{National Laboratory for Superlattices and Microstructures, Institute of Semiconductors, Chinese Academy of Sciences, P.O. Box 912, Beijing 100083, China}

\author{Zhi-Chuan Niu}
\affiliation{National Laboratory for Superlattices and Microstructures, Institute of Semiconductors, Chinese Academy of Sciences, P.O. Box 912, Beijing 100083, China}

\begin{abstract}

An upper bound between the information gain and state reversibility of weak measurement was first developed by Y. K. Cheong and S. W. Lee
[$\emph{Phys. Rev. Lett.}$ $\textbf{109}$, 150402 (2012)]. Their results are valid for arbitrary $\emph{d-level}$ quantum systems. In light of the commonly used qubit system in quantum information, a sharp tradeoff relation can be obtained. In this letter, this tradeoff relation is experimentally verified with polarization encoded single photons from a quantum dot. Furthermore, a complete traversal of weak measurement operators is realized, and the mapping to the least upper bound of this tradeoff relation is obtained. Our results complement the theoretical work and provide a universal ruler for the characterization of weak measurements.
\end{abstract}
 \maketitle

The tradeoff relation between the information gain and state disturbance\cite{Sciarrino} is a fundamental problem of quantum mechanics exemplified by the the famous Heisenberg uncertainty principle\cite{Kennard}, and it has important applications in quantum communication (quantum cryptography). It is widely believed that the stronger the measurement is, the more information can be extracted but the less coherence can be preserved in the system. The quantum state can be precisely estimated through  projective von Neumann measurements (PVNM)\cite{Heisenberg,Banaszek,Devetak,Ariano,Andersen,Sacchi}, n which the state is completely and irreversibly perturbed.

However, another type of measurement, known as a weak measurement, was first developed by Aharonov $\emph{et al.}$, has attracted much attention for its fundamental and enormous practical relevance, in the realm of quantum measurements \cite{Vaidman}. As a result of the weak interactions between the system and measurement apparatus, the quantum state collapses with limited and incomplete information extracted from the measurement; in other words, the coherence is partially preserved in the system rather than completely disturbed, as the case in PVNM. This remaining coherence can lead to a novel phenomenon in which the so-called weak values can exceed the range of eigenvalues associated with the observable in question. This enhancement effect can be used to precisely character
displacements requiring measurement sensitivities below the wavelength level \cite{Ritchie,Hosten}.

In weak measurements, as a result of the undisturbed coherence preserved in the system \cite{Ueda,Royer,Koashi}, the quantum state can be exactly retrieved with some probability \cite{Korotkov,Katz}. Therefore, a complementary relation should also exist between the information gain and state reversibility (which can be viewed as a measurement of disturbance). Using this complementary relation, joint observables can be measured through sequential measurements. Only when these measurements are weak, past events can be inferred about after their occurrence as a probabilistic wave collapse; thus, the evolution of the system can be characterized \cite{Mitchison}. With this method, Hardy's paradox can be well explained experimentally \cite{Lucien1,Lucien2,Lundeen,Yokota}. Similarly, an experimental violation of Heisenberg¡¯s measurement-disturbance relation is observed \cite{Ozawa,Rozema}. Y. S. Kim $\emph{et al.}$ demonstrated a scheme to protect entanglement from amplitude-damping decoherence using weak measurement and quantum measurement reversal\cite{Kim}. For the types of experiments that include joint observables, it is important to select the measurement strength, which in turn determines the final consequence. As a result, a quantitative description of this complementarity can provide a useful tool with which the most appropriate measurement strength can be decided. Many theoretical efforts have been devoted to quantitative description of this complementarity in weak measurements \cite{Nielsen,Braunstein,Buscemi,Luo,Terashima}. Recently, a clear expression of this relation was given by Y. K. Cheong and S. W. Lee \cite{Cheong}. Their results are valid for arbitrary $\emph{d-level}$ quantum systems and can be simplified to a sharp tradeoff relation for qubits, which can be expressed as
\begin{equation}
6G_{max}+P_{rev}=4
\end{equation}
where $G_{max}$ and $P_{rev}$ are the maximal value of mean estimation fidelity and reversibility after the weak measurement, respectively. Note that they are scaled in the range 0$\leq$$P_{rev}$$\leq$1 and 1/2$\leq$$G_{max}$$\leq$2/3.

The definitions of $G_{max}$ and $P_{rev}$ are consistent with those in Ref. [30], which are calculated as
\begin{equation}
G_{max}=\int d\varphi \sum_{r=1}^{2}p(r,|\varphi\rangle)|\langle\tilde{\varphi}_{r}|\varphi\rangle|^{2}
\end{equation}

\begin{equation}
P_{rev}=\sum_{r=1}^{2}|\langle\varphi|\hat{R}^{(r)}|\varphi\rangle_{r}|^{2}p(r,|\varphi\rangle)
\end{equation}
where $p(r,|\varphi\rangle)$ denotes the probability that the weak measurement outcome is $\emph{r}$, and the post-measurement state is $|\varphi\rangle_{r}$. $\tilde{\varphi}_{r}$ is the corresponding guessed state. ${R}^{(r)}$ is the reverse operator of the weak measurement operator $\hat{A}_{r}$.

This relation is the first direct
and quantitative link between information gain and reversibility, and it can be used as a basic method to characterize weak measurements potentially used in
quantum information processing. In addition, this relation is experimentally feasible because both the estimation fidelity and reversal probability are measurable quantities. To the best of our knowledge, this tradeoff relation has never been experimentally tested. Because its derivation process is quite mathematical, an experimental verification of this relation is fairly meaningful. Rather than the two single examples in Ref. [30], we describe both the initial incoming states and the weak measurement operator sets. As a result, a one-to-one mapping of the weak measurement operator to the tradeoff relation is given by our results. The final results agree well with the theoretical tradeoff relation. Our results are useful tools that assist in the determination of the measurement strength in quantum information processes.

Considering a theoretical qubit system $\varphi=\sqrt{\alpha}|H\rangle+e^{i\phi}\sqrt{\beta}|V\rangle$ (H and V denote the horizontal and vertical polarized states, respectively), a weak measurement acting on this state can be described by two operators written as
\begin{subequations}
\begin{equation}
\hat{A}_{1}=\sqrt{1-\varepsilon}|H\rangle\langle H|+\sqrt{1-\eta}|V\rangle\langle V|
\end{equation}
\begin{equation}
\hat{A}_{2}=\sqrt{\varepsilon}|H\rangle\langle H|+\sqrt{\eta}|V\rangle\langle V|
\end{equation}
\end{subequations}
where the range of $\varepsilon$ and $\eta$ are both from 0 to 1.

The adopted guessing strategy for the weak measurement is consistent with that in Ref. \cite{Kim}, which has been proved to be optimal to achieve the best information gain. The strategy for the state reversal is similar with the state recovery process through optical spin-echo effect \cite{xu}. As a result, for qubit, the reversal operators can be obtained through eigenstates flipping in weak measurement operators $\hat{A}_{r}$. With these strategies, the calculated results are
\begin{equation}
G_{max}=\frac{1}{6}(3+|\eta-\varepsilon|)
\end{equation}
\begin{equation}
P_{rev}=1-\varepsilon-\eta+2\varepsilon\eta
\end{equation}


From these results, it can be concluded that both $G_{max}$ and $P_{rev}$ are determined by the weak measurement assignments. Considering that the ranges of $\varepsilon$ and $\eta$ are both from 0 to 1, the scaled ranges of $G_{max}$ and $P_{rev}$ are consistant with Ref. [30].

 To experimentally realize a weak measurement, we adopt a Sagnac interferometer (SI) based on a polarized beam splitter (PBS), with a half wave plate (HWP) on either arm. The weak measurement parameters $\varepsilon$ and $\eta$ are solely decided by the angles of the HWPs. In our proposal, the state reversal process is also achieved through a similar SI apparatus. The entire experimental setup is shown in Fig. 1. A single InAs/GaAs quantum dot placed in cryostat is used as the single photon source (SPS) is the same as that used in Ref. \cite{Tang}, and the initial incoming states $\varphi$ of the emitted photons are prepared by the polarizer and HWP mounted in the programmable motor-driven rotation stage (RS1). From the above calculations, the final results are solely determined by the probability amplitude, hence the relative phase item $\phi$ does not play a role and can be omitted. As a result of the normalization, $\alpha$ is the unique variable that describes the initial incoming states, and the traversal of states can be realized through a rotation of RS1. The theoretical calculations contain an integration over the incoming states, which is achieved through a state traversal process in this experiment. In detail, 51 points of RS1 rotation angles are selected and the corresponding prepared states are $\varphi$=$\sqrt{0.02i}$$|H\rangle$+$\sqrt{1-0.02i}$$|V\rangle$(i=0,...,50). The first SI implements the weak measurement. Assuming that for $\hat{A}_{1}$, the angles of the two inserted HWPs mounted in RS2 and RS3 are set to be $\emph{a}$ and $\emph{b}$ (0$\leq$$\emph{a}$$\leq$$\pi/4$, $\pi/4$$\leq$$\emph{b}$$\leq$$\pi/2$), $\varepsilon$ and $\eta$ in Eq. (2) are calculated as $\sin^{2}(2a)$ and $\sin^{2}(2b)$, respectively. The other complementary operator, $\hat{A}_{2}$, can be executed when RS2 and RS3 are set to be $\pi/4-a$ and $3\pi/4-b$, respectively.

\begin{figure}[tbph]
\centering
\includegraphics[width=3.5in]{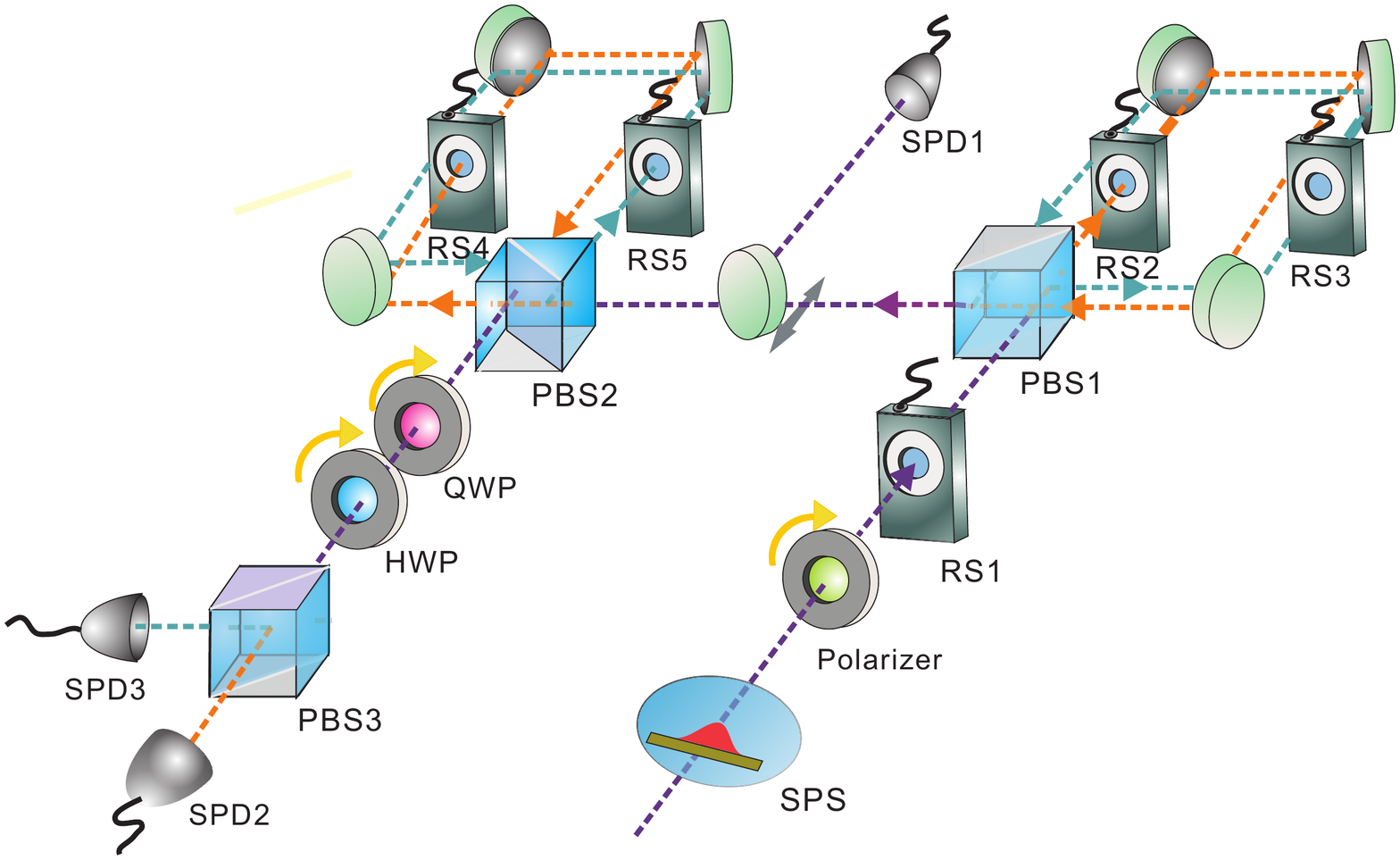}
\caption{\label{Fig1} The experimental setup. Single photons from the SPS are prepared to the initially measured linear polarized states. The weak measurement and reversal process are both implemented by SIs, in which a rotated HWP is inserted on either arm. The final tomography process of the reversed state is executed by a standard polarization analyzer, which involves a quarter wave plate (QWP), a HWP and a PBS. The outgoing photons from the interferometers are counted by single photon detectors (SPD). }
\end{figure}

If a silver flip mirror is inserted between the two SIs,
the outgoing photons from the first SI are counted by SPD1 and are denoted as $\emph{$C^{M}_{i}$}$(a,b) for the $i_{th}$ incoming state.
The experimentally measured value of $G_{max}$ is
\begin{equation}
\begin{split}
G_{max}= \\
\frac{1}{51}\sum_{i=0}^{50}\frac{\zeta C^{M}_{i}(a,b)+(1-\zeta)C^{M}_{i}(\pi/4-a,3\pi/4-b)}{C^{M}_{i}(a,b)+C^{M}_{i}(\pi/4-a,3\pi/4-b)}
\end{split}
\end{equation}
where $\zeta$ is the estimation fidelity and can be calculated as
\begin{equation}
\zeta =
\begin{cases}
0.02i,  & \mbox{if }\mbox{$\sin(2a)$ $<$ $\sin(2b)$} \\
1-0.02i, & \mbox{if }\mbox{$\sin(2a)$ $>$ $\sin(2b)$}
\end{cases}
\end{equation}

According to the original definition of $\hat{R}^{(r)}$, it is the exclusive reversal operator corresponding to $\hat{A}_{r}$, and it can reverse $|\varphi\rangle_{r}$ exactly back to the initial income state. The reversal fidelity is theoretically equal to 1, with a probability less than 1. This state reversal process is realized by the second SI apparatus, which is approximately identical to the first one except that the angle setting for the HWPs mounted in RS4 and RS5 are exchanged between the two arms. In other words, the corresponding angle settings in the second SI are ($\emph{b}$,$\emph{a}$) and ($3\pi/4-b$,$\pi/4-a$) for $\hat{R}^{(1)}$ and $\hat{R}^{(2)}$, respectively. With these assignments, the photons emitted from the crossing output port are reversed to the initial incoming states. A polarization analyzer implements the state tomography of the output photons. $C^{R}_{i}(b,a)$ denotes the summation of photon counts recorded by SPD2 and SPD3, when the angles of RS4 and RS5 are set at $\emph{b}$ and $\emph{a}$.

The reversibility $P_{rev}$ can be calculated as
\begin{equation}
P_{rev}=\frac{1}{51}\sum_{i=0}^{50}\frac{C^{R}_{i}(b,a)+C^{R}_{i}(3\pi/4-b,\pi/4-a)}{C^{M}_{i}(a,b)+C^{M}_{i}(\pi/4-a,3\pi/4-b)}
\end{equation}

The definition of $G_{max}$ is independent of the initial measured state of the photons, because an integration is executed over all states. In the experiment, this integration process is realized by averaging over the 51 uniformly distributed states. Referring to the individual incoming states with the measurement apparatus unaltered, the quantity of the extracted information is state dependent while the reversibility is a constant. The experimental results for $\varepsilon=0.25$ and $\eta=0.75$ are shown in Fig. 2. The errors of the results from this experiment mainly come from the random fluctuation of the photon counts and the uncertainties in aligning the wave plates. The error bars are estimated by the square root of the variance of 50 measurement values. The state independence of the reversibility results from the symmetrical assignments of the two SIs. Note that the quantity of extracted information can exceed 2/3 for part of the incoming states, but the average over these states is still within the range 1/2$\leq$$G_{max}$$\leq$2/3.

\begin{figure}[tbph]
\centering
\includegraphics[width=3.5in]{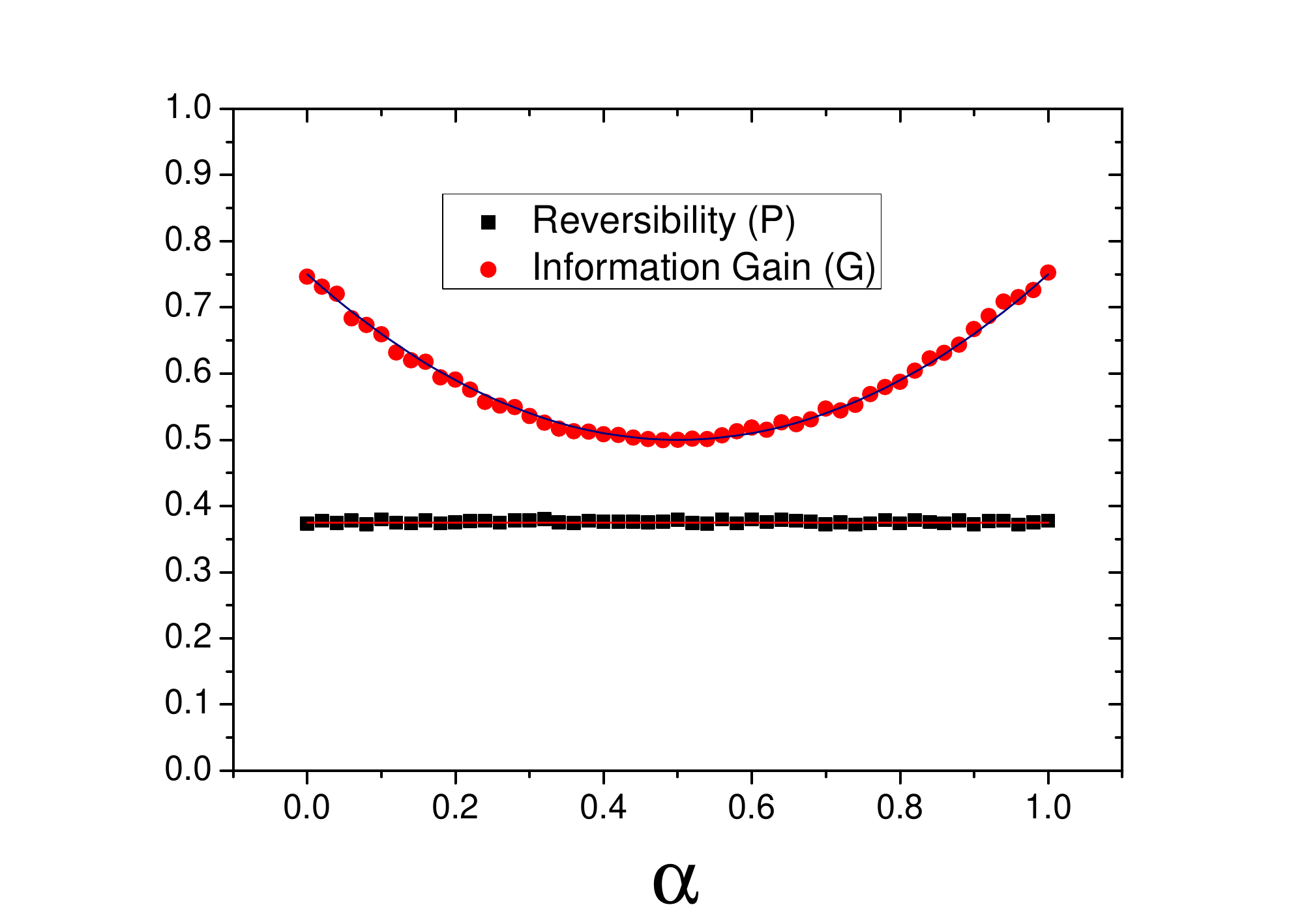}
\caption{\label{Fig2} The initial state traversal results when the measurement apparatus is set to make $\varepsilon=0.25$ and $\eta=0.75$. The sample points of $\alpha$ are uniformly selected with an interval of 0.02. The distributions of both information quantity and reversibility are shown in the figure. The solid lines are the theoretical, see the text for detail. Error bars not shown are smaller than the marker dimensions.}
\end{figure}

In our experiment, we measure 6$G_{max}+P_{rev}$ for each weak measurement determined by the values of $\emph{a}$ and $\emph{b}$. A two-dimensional traversal of $\emph{a}$ and $\emph{b}$ is realized by rotating RS2 and RS3. The measured points are uniformly selected for both $\varepsilon$ and $\eta$ in the traversal process. For certain value of $\varepsilon$, 16 values of $\eta$ are measured ranging from 0 to 1 and vice versa. Totally 256 points were selected in the experiment, corresponding to 256 measurement operator sets. The interpolating three-dimensional mapping of measurement operator sets to the value of 6$G_{max}+P_{rev}$ is shown in Fig. 3, accompanied by the theoretical result calculated from Eq. (8) and (9).
Both sets of results show a tapered distribution with the four sides equal to 4 and the internal points gradually decreasing from the sides to the center. The upper bound is 4 for qubits, which is consistent with the theoretical result. For weak measurements on the four sides, the tradeoff relation described by Eq. (1) is strictly satisfied. The bound approaches the minimum value of 3.5 at the center point, where $\varepsilon$ = $\eta$ = 0.5. Actually, for all points on the diagonal line with $\varepsilon$ = $\eta$$\neq$ 0 or 1, the operation of $\hat{A}_{r}$ can be regarded as a beam splitter, with which no measurement acts on the incoming photons. The reversibility of these points is no longer in accordance with the definition in Ref. [30], and hence should be excluded from the results. The two corner points where $|$$\varepsilon$ - $\eta$$|$ = 1 correspond to the PVNM, and reversibility equal to 0 and the best estimation fidelity are achieved at either side. This result shows that the von Neumann measurement can not be reversed in any case; in other words, the extracted information can not be erased.
\begin{figure}[tbph]
\centering
\includegraphics[width=3.5in]{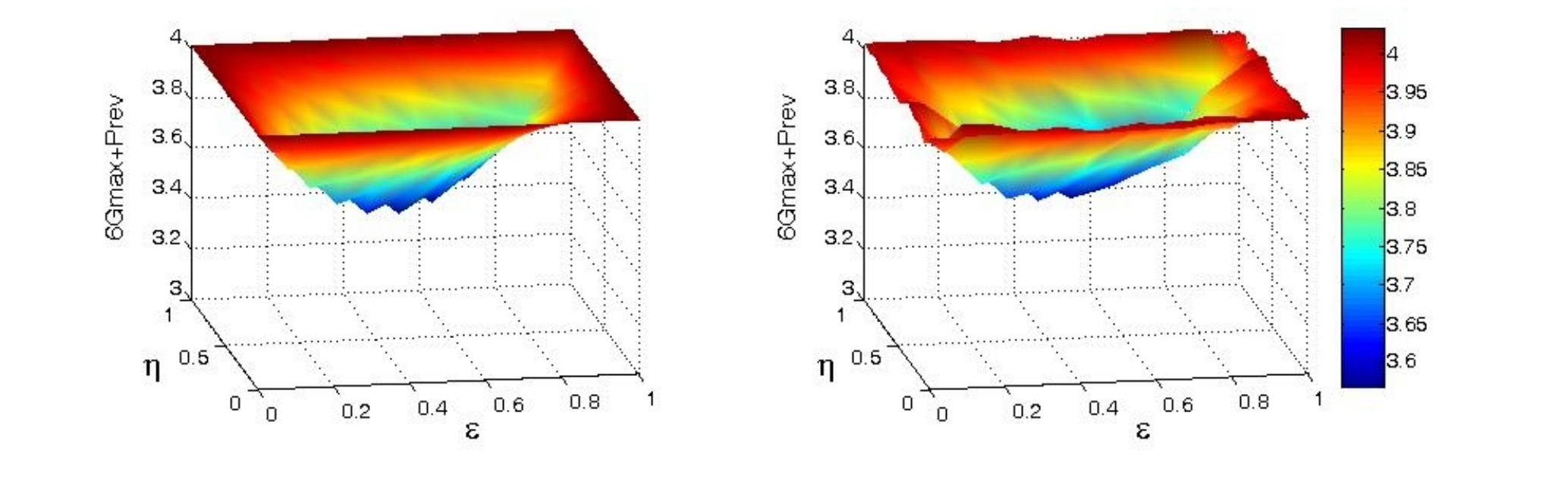}
\caption{\label{Fig3} (a) Theoretical and (b) experimental mapping of balance characteristic to two-dimensional traversal of $\varepsilon$ and $\eta$.}
\end{figure}
To test that the photons emitting from the second SI have been precisely reversed, quantum tomography is executed on a set of initial incoming states with the identical measurement apparatus. For $\varepsilon=0.25$ and $\eta=0.75$, the fidelities of the reversal states for the 51 initial incoming states are shown in Fig. 4(a). For all of these states, the reversed fidelities exceed 0.99, indicating a successful reversal operation in the experiment. These high fidelities rely on the excellent quality of the two SIs, which both exhibited the extinction ratios greater than 1000:1 throughout the data recording period.

The measurement strength is closely related to the relative value of $\varepsilon$ and $\eta$. As discussed above, when $\varepsilon$ = $\eta$, the measurement apparatus is equivalent to a beam splitter and the measurement strength can be regarded as 0, while $|$$\varepsilon$ - $\eta$$|$ = $\emph{1}$ would indicate a PVNM measurement. To further clarify this point, we remove the $\varepsilon=0$ cross section, and show the dependency of 6$G_{max}+P_{rev}$, 6$G_{max}$ and $P_{rev}$ on the value of $\eta$ in Fig. 4(b). A strict tradeoff relation is observed from this picture. When $\eta$ grows from 0 to 1, the measurement strength increases, resulting in a linear ascension of 6$G_{max}$ and a linearly descending of $P_{rev}$. The identically varying rates keep 6$G_{max}+P_{rev}$ unaltered in this process. For the origin point, no measurement is executed. $P_{rev}$=1 indicates that the state is perfectly preserved while $G_{max}$ achieves the minimum of 0.5, which is equivalent to a random bet about the state. Conversely, a PVNM leads to a maximum value of $G_{max}$ of 2/3, and the state is completely disturbed with $P_{rev}$=0.

\begin{figure}[tbph]
\centering
\includegraphics[width=3.5in]{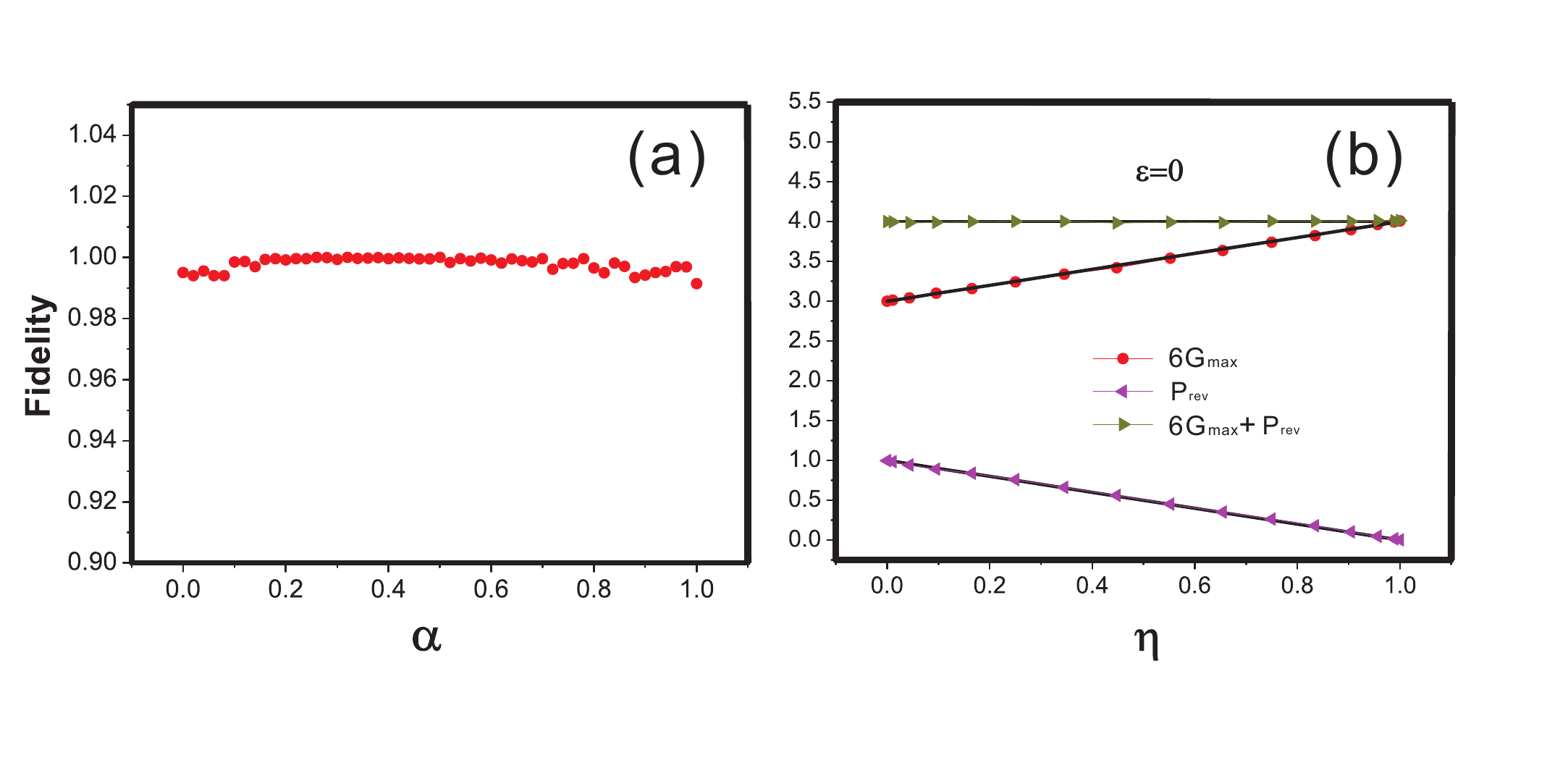}
\caption{\label{Fig4} (a) The fidelities of the reversed states when $\varepsilon=0.25$ and $\eta=0.75$. (b)The distributions of 6$G_{max}+P_{rev}$, 6$G_{max}$ and $P_{rev}$ with an $\varepsilon=0$ cross section when $\eta$ varies from 0 to 1. The solid lines are the theory according to Eq.
(8) and (9), see the text for details. Error bars not shown are smaller than the marker dimensions.}
\end{figure}

Corresponding to the complete loss of reversibility, the increasing amplitude of $G_{max}$ is only 1/6. In other words, the state can not be fully estimated even through PVNM with sufficiently large sample size. This independent PVNM is very different from the state tomography process, which is also based on PVNM. Consider a qubit state with no relative phase between the H and V bases, state tomography can be achieved simply through repeated PVNM on these two bases. The estimation fidelity can be infinitesimally close to 1. However, for independent PVNM with the same experimental setup, the average estimation fidelity can not exceed 2/3. Actually, the information extracted from a single PVNM operation is the same for these two measurement processes. The factor that is different is that, the independent PVNM makes an estimation about the state based on this partial information, which inevitably induces errors and dilutes the extracted information. Unlike the independent PVNM, state estimation through tomography is based on the summation of information extracted from numerous times of PVNM, and the final error induced can be arbitrarily small.

In summary, we experimentally tested the tradeoff relation concering the balance between information gain and state reversibility in weak measurement.
The experimental results are strongly consistent with the conclusion reached by Y. K. Cheong and S. W. Lee. In addition, a 3-dimensional mapping is given in this paper with which this balance can be quantitatively described for arbitrary weak measurement. Further analysis reveals the dependency of estimation fidelity and state reversibility on the measurement strength. Our results extend the theoretical work and provide a universal ruler for the characterization of weak measurements.

This work was supported by the National Basic Research Program, the National Natural Science Foundation of China (Grants No. 60921091, No. 10874162, No. 90921015, No.11105135 ), the Fundamental Research Funds for the Central Universities (No. WK2470000011, WK2470000004, WK2470000006, WJ2470000007), and the 973 program(2013CB933304).

\end{document}